\documentstyle[aps,graphicx,prl,multicol]{revtex}

\begin{document}

\title{On a dual model for 5$f$ electrons}
\author{D.V.~Efremov, N.~Hasselmann, E.~Runge, P.~Fulde}
\address{%
Max-Planck-Institut f\"ur Physik komplexer Systeme, 
N\"othnitzer Str. 38, 01187 Dresden, Germany}
\author{G. Zwicknagl}
\address{%
Institut f\"ur Mathematische Physik, 
Technische Universit\"at Braunschweig, Mendelssohnstr. 3, 
38106 Braunschweig, Germany}
\date{\today}
\maketitle
\begin{abstract}
We study the effect of intra-atomic correlations on anisotropies of 
hopping of 5$f$ electrons. It is shown that they may considerably 
enhance these anisotropies to the extend that electrons in some of the 
5$f$ orbitals  remain localized. This provide a microscopic  basis for a 
previously made assumption that some of the 5$f$ electrons 
must be partially localized when Fermi surfaces and effective masses of the 
component are 
calculated. Calculations are performed on two- and three-sites clusters. They 
include  phase  diagrams in the presence and absence of an external 
magnetic field as function of anisotropic hopping. 

\end{abstract}
\begin{multicols}{1}
\narrowtext
\section{Introduction}

Systems with strongly correlated electrons are a subject of considerable
research efforts. Among those are the ones with 5$f$ electrons, i.~e.,
involving actinide ions of particular interest. While 4$f$ electrons, because
of their closeness to the nuclei, are correlated as strongly as possible on an
atom, the 5$f$ electrons have a larger spatial extent and tend more
towards delocalization. The degree of delocalization is even more pronounced
when 3$d$ electrons are considered. By continuing this line of arguments one
finds that $s$ electrons are the least correlated ones on an atom and
have the largest tendency to delocalize. 

Returning to 5$f$ electrons, the question arises 
under which circumstances localized features dominate delocalized ones
or vice versa. In particular we want to follow up the suggestion which has
been made previously \cite{GYF02} that 5$f$ electrons have a 
dual character in the sense that
some of the 5$f$ orbitals lead to delocalization when occupied by
electrons while others result in localization. It was demonstrated
that a division of 5$f$ orbitals
into delocalized and localized ones is able to explain  the heavy 
quasiparticles observed in UPt$_3$ and UPd$_2$Al$_3$ including their
mass anisotropies \cite{GYF02,GEZJapan}. 
However, the question remained unanswered as regards 
the microscopic origin of this division of orbitals. Apparently, strong
intra-atomic correlations seem to enhance anisotropies in the kinetic energy
matrix elements which reflect anisotropies in the hybridization with the
chemical environment.

The purpose of the present work is to clarify the physical origin of the
dual character of 5$f$ electrons. For that purpose clusters of two and
three sites are considered. Because of the large spin-orbit coupling 
each 5$f$ electron is characterized by its total angular momentum $j=5/2$
and azimuthal quantum number $j_z$. We do not include $j=7/2$ states in
our analysis since their energy is higher by approximately 1eV \cite{albers}.
We shall treat clusters where the average number of 5$f$ electrons per site
is either 2.5 or 2.67, values which are
close to the 5$f$ occupancy of U ions in compounds like UPt$_3$ \cite{albers},
UPd$_2$Al$_3$ or UGe$_2$ \cite{Kernavanois,Shick}. 

Depending on $j_z$ different hopping matrix elements
are assumed for the $f$ electrons in the cluster. 
The key result of the present investigation is to demonstrate
that indeed anisotropies in the hopping are strongly enhanced by
intra-atomic correlations.

The paper is organized as follows. After defining the model Hamiltonian
a two-site and a three-site cluster are investigated. Various properties are
calculated, like the ground-state energy and magnetization as well as the
density of states and its magnetic field dependence. This is done for different
ratios, of hopping. Partial 
localization is the subject of the subsequent Section \ref{secPartial}. A
discussion and the conclusions are found in Sect. VI. The Appendices are
devoted to perturbation theory, the atomic limit of vanishing inter-atomic
electron transfer and to a more detailed analysis of the ground state's
composition. 

\section{Model Hamiltonian}
We investigate a microscopic model of 5$f$ electrons which includes
local Coulomb interactions and hopping terms between nearest neighbor sites.
The model Hamiltonian is  
\begin{eqnarray}
H & = & - \sum _{\langle nm\rangle, j_{z}}\, t_{j_{z}}\left( 
  c_{j_{z}}^{\dagger}(n)\,c_{j_{z}}(m)+h.c.\right) \nonumber\\
  &  & -\sum _{n,j_{z}}\, h_{n}\, j_{z}c_{j_{z}}^{\dagger }(n)
  \,c_{j_{z}}(n)+
H_{\mathrm{Coul}}
\label{eq:TwoSiteHamiltonian}
\end{eqnarray}
where the first sum is over neighboring sites $\langle nm\rangle$. 
Furthermore $c_{j_{z}}^{\dagger}(n)$ ($c_{j_{z}}(n)$), creates (annihilates) an
electron at site $n$ in the $5f\ j = 5/2$ state with
$j_z=-5/2,\dots,5/2$. We will
consider two and three sites models. 
The effective hopping between sites results from
the hybridization of the 5$f$ states with the orbitals of the ligands
and depends generally on the crystal structure. Rather than trying to
exhaust all possible different lattice symmetries, we shall concentrate
here on the special case that hopping conserves $j_z$. While this is
certainly an idealization, it allows us to concentrate on our main
interest, i.~e., a study of
the influence of atomic correlations on the renormalization of
hybridization matrix elements.
The parameters $t_{j_z}(=t_{-j_z})$ are chosen 
in accordance with density-functional calculations for bulk material
which use $jj_z$ basis states. 
We also include local
magnetic fields $h(n)$ to account for the influence of
ferromagnetic (FM) or antiferromagnetic (AFM) 
internal fields.
The local Coulomb interactions can be written in the form
\begin{eqnarray}
\label{eqCoulomb}
  H_{\mathrm{Coul}}=&& \frac{1}{2}
  \sum_{n} 
  \sum_{j_{z1},\dots,j_{z4}} \nonumber \\ &&
  U_{j_{z1}j_{z2},j_{z3}j_{z4}}
  c_{j_{z1}}^{\dagger }(n)\,c_{j_{z2}}^{\dagger }(n)\,c_{j_{z3}}(n)\,
  c_{j_{z4}}(n)
\end{eqnarray}
with Coulomb matrix elements 
\begin{eqnarray}
  \lefteqn{U_{j_{z1}j_{z2},j_{z3}j_{z4}}
  = \delta_{j_{z1}+j_{z2},j_{z3}+j_{z4}} \times} 
  & & \nonumber\\
  & & 
  \sum_J {\textstyle \langle \frac{5}{2}\,\frac{5}{2}\,j_{z1}\,j_{z2} | 
                     JJ_{z} \rangle }
  U_{J}  {\textstyle \langle JJ_{z}| \frac{5}{2}\,\frac{5}{2}
           \,j_{z3}\,j_{z4} \rangle }.
\end{eqnarray}
Here $J$ denotes the total angular momentum of two electrons and 
$J_{z}=j_{z1}+j_{z2}=j_{z3}+j_{z4} $. The sum is restricted 
by the antisymmetry of the Clebsch-Gordan coefficients 
\( \langle \frac{5}{2}\,\frac{5}{2}\,j_{z1}\,j_{z2}| JJ_{z}\rangle  \)
to even
values $ J=0,2,4 $. 
We use in the actual calculations \( U_{J} \) values 
which are determined from LDA wavefunctions for 
UPt\( _{3} \)\cite{GYF02}, 
i.~e.,~$U_{J=4}=17.21 eV$, 
$U_{J=2}=18.28 eV$, and $U_{J=0}=21.00 eV$.
We expect $U_{J=4}<U_{J=2}<U_{J=0}$
to always hold for Coulomb interactions, independently of the
chemical environment.
In contrast,
the relative order of the hopping matrix elements
will vary strongly from one compound to the next. 

The average Coulomb repulsion of about 20\,eV
is irrelevant for the low-energy physics of the model. 
It simply restricts the relevant configurations to states so that
the sites are occupied either by 2 or 3 electrons.
The low-energy sector is solely determined by the differences of the 
$U_J$ values, which are of the order of 1\,eV and thus slightly larger
than typical bare $f$~bandwidths. The latter are obtained, e.g., from LDA
calculations for metallic uranium compounds like UPt$_3$. Note that restricting
the model to $f^2$ and $f^3$ configurations is equivalent to let
the different $U_J\to \infty$ while their differences remain finite.

To mimic the situation in the U-based heavy-fermion compounds we consider
the intermediate valence regime. 
Note that in the absence of a magnetic field
all states of the two-site model with 5 electrons  
will be at least doubly 
degenerate because of Kramers' degeneracy.

The Hamiltonian Eq.~(\ref{eq:TwoSiteHamiltonian}) conserves 
 ${\cal J}_z=\sum_n J_z(n)$ where ${\cal J}_z$ is the z-component of the total 
angular momentum of the system and the
$J_z(n)$ refer to angular momentum projections
on individual sites. We shall therefore characterize the
eigenstates by their ${\cal J}_z$ value. 
For isotropic hopping
$t_{j_z}=t=\mbox{\rm const}$ the system is rotationally invariant. Then
$\mbox{\boldmath $\bf \cal J$}^2$ provides an additional good quantum number.

\section{Two site model}
As mentioned before it should suffice to limit ourselves to the
subspace of $f^2$ and $f^3$ configurations, because of the large
values of $U_J$.
To assess the validity of this
simplification we shall compare the predicted ground states for the $f^2$-$f^3$
restricted model with those
obtained by diagonalizing the Hamiltonian 
Eq.~(\ref{eq:TwoSiteHamiltonian}) in the full space.
The rational for the restriction to $f^2$-$f^3$ is not the
reduction of the numerical diagonalization effort
by a factor of about two, but the conceptual simplicity gained 
for the interpretation and presentation of the results
for, e.g., local spin configurations and orbital occupations.
In the full model, interactions of the order of ${\bar{t}}^2/\bar{U}$
are present which arise
from virtual $ f^2$-$f^3 \leftrightarrow f^1$-$f^4$
transitions 
($\bar{t}$ and $\bar{U}$ are representative values
of $t_{j_z}$ and $U_J$, respectively.)
The restricted model ceases to be a good approximation 
when these interactions 
become comparable to the bare parameters of the model. 

For most of the calculations below we shall assume
\( t_{5/2}=t_{1/2} \) and \( t_{3/2}\)
as independent parameters, a choice motivated by LDA calculations
for UPt$_3$ \cite{GYF02}
where $t_{3/2}>t_{1/2}$, $t_{5/2}$. However, partial
localization, which we will discuss in Section \ref{secPartial}, is
also found for other parameter choices, e.~g., by choosing
\( t_{5/2}=t_{3/2} \) and \( t_{1/2}\) as independent parameters. The
origin is qualitatively identical for both choices, as we will
demonstrate later. The results presented 
below  are obtained from numerical studies
of the Hamiltonian (\ref{eq:TwoSiteHamiltonian}).

\subsection{Ground-state energy}
\begin{figure}[]
\begin{center}
\noindent 
\includegraphics[width=0.9\columnwidth,angle=0]{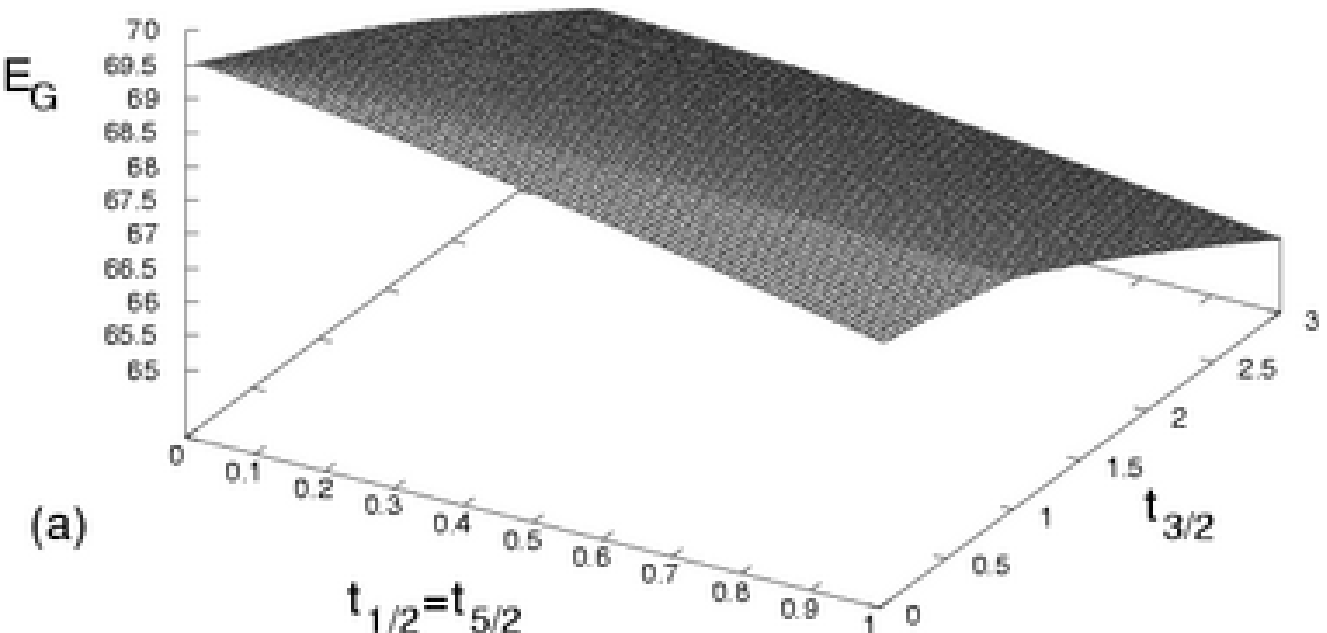} 
\newline \noindent
\includegraphics[width=0.9\columnwidth,angle=0]{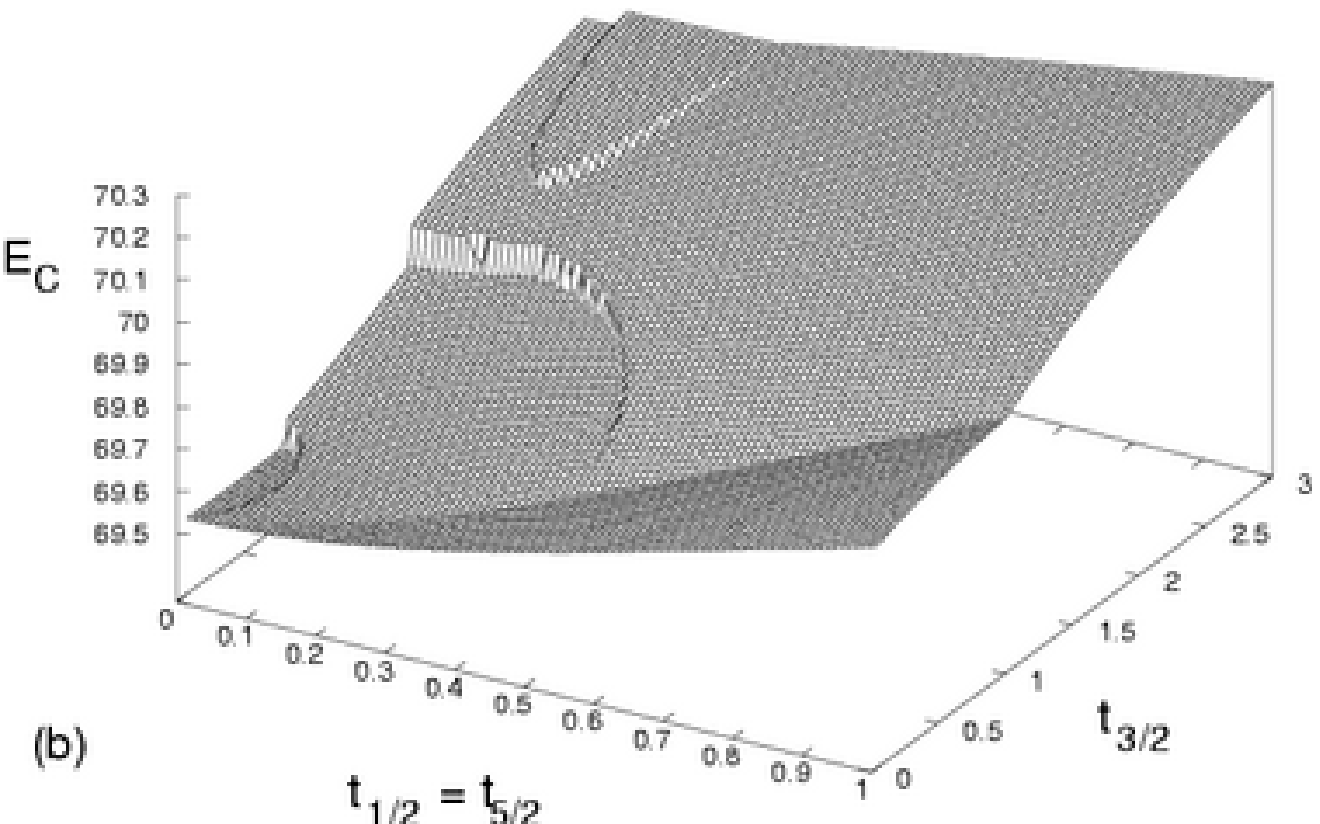} 
\end{center}
\caption{(a) Total ground-state energy as a function of
$t_{3/2}$ and $t_{5/2}=t_{1/2}$. 
(b) Expectation value of the Coulomb interaction 
  $\langle \Psi_{\mathrm{gs}} |H_{\mathrm{Coul}}| 
   \Psi_{\mathrm{gs}} \rangle $  
in the ground state $\Psi_{\mathrm{gs}}$ 
vs.\ hopping parameters $t_{3/2}$ and $t_{5/2}=t_{1/2}$. 
}\label{fig:ThreeEnergies}
\end{figure}

Figure~\ref{fig:ThreeEnergies}a shows the variation 
of the ground-state energy $E_G$ as a function of the
hopping parameters  $t_{3/2}$ and $t_{5/2}=t_{1/2}$. 
This energy  is smooth except for a kink along 
the isotropic line $t_{1/2}=t_{3/2}=t_{5/2}$. 

Level crossings 
are apparent from discontinuous jumps of 
the expectation value $E_C$ 
of the Coulomb interaction with respect to the ground state, see 
~Fig.~\ref{fig:ThreeEnergies}b. 
The ground-state energy at vanishing hopping 
is 69.53 eV which is the
sum of the Coulomb energy of the $J=4$ $f^2$
configuration and the $J=9/2$ $f^3$ configuration,
\begin{eqnarray} 
E_G(\{t_{j_z}=0\}) &=& E_C(f^2,J=4)+E_C(f^3,J=9/2) \nonumber \\ &=&
\left( 3+\frac{5}{14} \right)U_{J=4} +  \frac{9}{14} U_{J=2} \ .
\end{eqnarray}
In general, the ground-state energy decreases 
with increasing hopping due to a lowering of
the kinetic energy. This is however 
partially compensated for by an 
increase in the interaction energy. 
The jumps indicate quantum phase transitions,
which result from the competition between Coulomb
interactions and anisotropic hopping or hybridization.

\subsection{Magnetization}\label{subSecMagnGS}
\begin{figure}[h t b]
\begin{center}
\includegraphics[width=0.8\columnwidth]{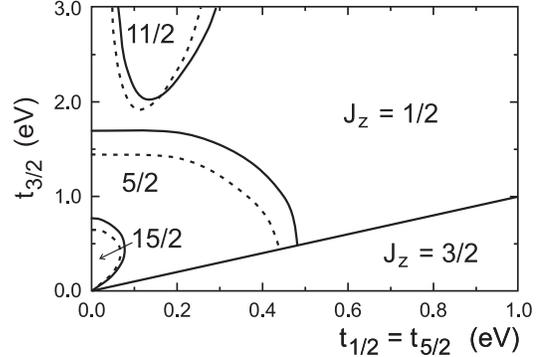}
\end{center}
\caption{
  Phase diagram in the  $t_{1/2}$(=$t_{5/2}$) vs.\ $t_{3/2}$ plane
  derived from the total magnetization ${\cal J}_z$ of the ground state.
  Solid lines: for the $f^2$-$f^3$ model.
  Dashed lines: with no restriction on the occupancy.
}\label{f2f3_PhasenTT2.eps}
\end{figure}

In order to characterize the ground state of the two-site cluster
for different choices of the hopping matrix elements $t_{j_z}$ we study the
total magnetization characterized by ${\cal J}_z$ in the presence of a vanishingly
small field. Results are shown in Fig.~\ref{f2f3_PhasenTT2.eps}.
The different ground states are separated by solid lines when we allow
for $f^2$ and $f^3$ configurations only. 
Dashed lines show the ground-state
boundaries when this restriction is lifted. The later are 
obtained from diagonalization of $H$ 
in the full space of 5-electron states. Because of the small
differences we shall limit ourselves to $f^2$-$f^3$ configurations only.
Five different phases are visible in Fig.~\ref{f2f3_PhasenTT2.eps}, 
i.e., two ``high-spin'' phases 
(${\cal J}_z=15/2$, ${\cal J}_z=11/2$) and three ``low-spin'' phases 
(${\cal J}_z=5/2,3/2,1/2$).
In  the ``high-spin'' phases which prevail for 
$t_{3/2} \gg t_{1/2}=t_{5/2}$ the two sites are ferromagnetically correlated,
$\big< {\bf J}(a)\cdot {\bf J}(b)\big> > 0$, 
while in the low-spin phases the correlations are antiferromagnetic (AFM),
$\big< {\bf J}(a)\cdot 
{\bf J}(b)\big> < 0)$. Here and below we denote
the two different sites by $a$ and $b$.

A special case is $t_{1/2}=t_{3/2}=t_{5/2}=t$ (isotropic case). Here
the ground state is always six-fold degenerate with ${\cal J}=5/2$. 
This is plausible in the limit of large values of $t$: all electrons
will occupy orbitals with different $j_z$ in order to maximize the kinetic
energy. With a total of five electrons one of the $j_z$ orbitals 
will remain empty, implying a six-fold degenerate ground state.

Even when $U_J \to \infty$, implying the $f^2$-$f^3$ model, 
but with a difference
$|\Delta U_J|$ of any pair of $U_J$ values much larger than $t$, i.e.,$|\Delta
U_J|/t \to 0$, one can see that ${\cal J}=5/2$. In that case the six-fold degenerate
ground states  can be explicitely written in the form   
\begin{eqnarray}
\lim_{|\Delta U_J|/t \to 0}  | \Psi_{j_z}\rangle = \sqrt{\textstyle\frac{8}{5}} 
  \,\,{\hat{\cal P}_{f^2\mbox{-}f^3}}
  \, \prod_{j_z'\not=j_z}\frac{ c^\dagger_{j_z'}(a)+ 
  c^\dagger_{j_z'}(b)}{\sqrt{2}} 
  \,|0\rangle 
  \,,
\label{corrwavefunction}
\end{eqnarray}
where ${\hat{\cal P}_{f^2\mbox{-}f^3}}$ is the projector allowing 
only for atomic $f^2$ and $f^3$ configurations.
One implication is that the
kinetic energy is only $-3t$ instead of
$-5 t$ as for free electrons, 
because at most three out of five electrons 
find free places to hop onto. Note that $j_{z} \neq j_{z'}$ in Eq. (\ref{corrwavefunction}). 
While for finite values of $t/|\Delta U_J|$ the ground-state 
wavefunctions become vastly 
more complex, 
they remain eigenfunctions of ${\cal J}=5/2$.
Simple results are found again in the limit $t\rightarrow 0$, 
see Appendix~\ref{appPerturbation}.

As one moves away from the isotropic line with ${\cal J}=5/2$ the degeneracy
of different ${\cal J}_z$ states is lifted.  Depending on $t_{3/2}$ and
$t_{1/2}=t_{5/2}$ one of them becomes the ground state while ${\cal J}$ is no
longer preserved (see Fig.~\ref{f2f3_PhasenTT2.eps}).

The phase with ${\cal J}_z=15/2$ 
appears when $t_{3/2}\gg t_{1/2}$($=t_{5/2}$). 
It corresponds to a very simple eigenstate in which all occupied orbitals
have $j_{z}>0$. One finds
\begin{equation}
|\Psi \rangle_{\frac{15}{2}} = 
\frac{c^\dagger_{3/2}(a)+c^\dagger_{3/2}(b) }{\sqrt{2}}\,
 c^\dagger_{5/2}(a)
 c^\dagger_{1/2}(a)
 c^\dagger_{5/2}(b)
 c^\dagger_{1/2}(b)\, | 0 \rangle
\label{eq:FerroHighState15}
\end{equation}
with kinetic energy $- t_{3/2}$. 
The $j_z$=5/2 and $j_z$=1/2 orbitals are occupied on both sites while
$j_z=3/2$ is  occupied with probability $1/2$ on each of the two sites.
Note that $|\Psi \rangle_{15/2}$ is separately an eigenstate of the Coulomb 
as well as  the  kinetic energy terms. 
While the Coulomb energy is at its absolute minimum, there are no 
energy contributions resulting from $t_{1/2}$ and $t_{3/2}$ due to the limitation
to $f^2$ and $f^3$ configurations. 
This can be considered as an extreme case of partial localization: 
the  $j_z$=5/2 and $j_z$=1/2 states are fully localized and the   
$j_z$=3/2 state is maximally delocalized.
When the hopping matrix elements increase,
a discontinuous jump takes place into a state with ${\cal J}_z=5/2$.
The corresponding increase in Coulomb energy is seen 
in Fig.~\ref{fig:ThreeEnergies}b.

The variation of the ground-state energy along the line $t_{1/2}=t_{5/2}=0$
can be qualitatively understood from the following argument: 
For sufficiently small hopping matrix elements $t_{3/2}$
 the high spin ${\cal J}_z=15/2$~state is lowest in energy
because it minimizes the Coulomb energy.
With increasing values of $t_{3/2}$ it becomes more favorable
to partially break up the local Hund correlations in order to gain kinetic
energy by partially populating both $j_z=3/2$ and the $j_z=-3/2$ orbitals.
In the limit of large  hopping, their angular momenta components cancel 
and ${\cal J}_z$ is given by the $j_z$ components of 
the  remaining electrons, i.e., by
$(\frac{5}{2}+\frac{1}{2})\pm \frac{5}{2}$ resulting in 
${\cal J}_z=11/2$ or 1/2. 
For intermediate hopping 
the total  angular momentum of states were 
both $\pm 3/2$ and $\pm 1/2$ are occupied is given
by the remaining $5/2$ orbitals. 
For more details on the occupation of the individual $j_z$ orbitals in the 
ground state  see  Appendix  \ref{appAnalysis}.

\begin{figure}[hbt]
\includegraphics[width=0.7\columnwidth]{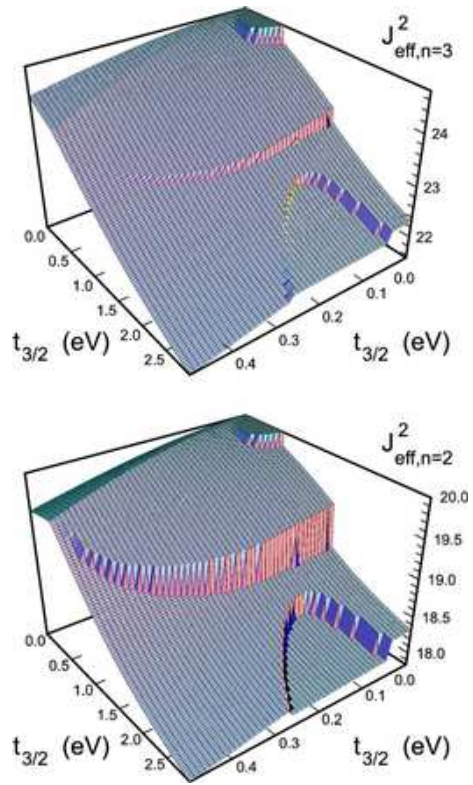}
\vspace{.5cm}
\caption{
  Total momentum as defined by Eq.~(\ref{eqDefQ2Q3})
  for sites with local $f^3$ (upper panel) and $f^2$
  configurations (lower panel), respectively. 
  Note the scale and the smallness of the deviation
  from the values $24.75 = ((9/2)+1)\cdot(9/2)$ and 
  $20 = (4+1)\cdot 4$ corresponding to fully polarized
   $f^3$ and $f^2$ configurations. 
}\label{figQ2Q3}
\end{figure}

While the total angular momentum ${\cal J}$ depends strongly on the
anisotropy, the local angular momenta change very little. 
Fig.~\ref{figQ2Q3} shows for different 
anisotropies the local spins defined 
in terms of projectors onto configurations
with given local $f$-occupation as
\begin{equation}
  \label{eqDefQ2Q3}
J^{\,2}_{{\mathrm eff},\mu=2(3)} = \sum_{n=a,b} 
  \big\langle {\bf J}(n)^{\,2} \,\hat{P}_{\mu=2(3)}(n)  
  \big\rangle
  \,.
\end{equation}

The projector $\hat{P}_\mu$ selects atomic configurations with $\mu$
electrons. 
The small range of $J^{\,2}_{{\mathrm eff},\mu}$ variation 
in Fig.~\ref{figQ2Q3} is remarkable. 
Hopping elements up to 0.5\,eV reduce the local moments 
 by less than 5\% relative to their maximal 
values 24.75 and 20 for $f^{2}$ and $f^{3}$ 
configurations, respectively. 
One observes a reduction of the local spin expectation
values at  large hopping 
for both the $f^{2}$ and the $f^{3}$ site.
A decrease of  $J_{\mathrm{eff},\mu=2(3)}$ implies 
for $U_{J=4} < U_{J=0,2}$, an increase
of the Coulomb energy. 
Thus, Figure~\ref{figQ2Q3} demonstrates the 
competition between Coulomb interaction and
hopping.

\subsection{Density of states}\label{subSecDOS}
In the absence of any hopping, i.e., for  
$t_{3/2}$\,=\,$t_{1/2}$\,=\,$t_{5/2}$\,=\,$0$, the ground state 
is 180-fold degenerate (see Appendix \ref{appPerturbation}).
At finite isotropic
hopping this degeneracy is only partially lifted. Then $J(a)$ and
$J(b)$ are
no longer separately conserved, but the total $\cal J$ still is.
\begin{figure}[htb]
\begin{center}
\includegraphics[width=1.0\columnwidth]{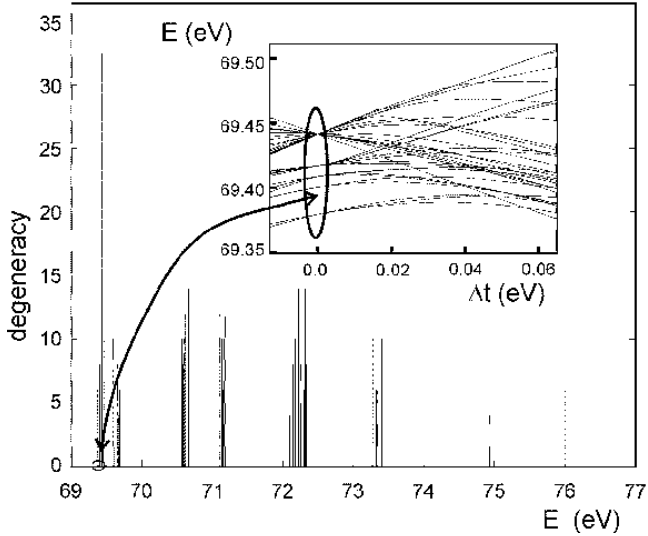}
\end{center}
\caption{
Energy levels and their degeneracies for isotropic hopping, 
$t_{1/2}=t_{5/2}=t_{3/2}=t=0.1$\,eV.
Inset: Anisotropic hopping $\Delta t=t_{3/2}-t_{1/2}$ 
lifts the high degeneracies.
\\
}\label{fig:dosSplitting}
\end{figure}
 
Thus, along the isotropic line, spherical symmetry still leads
to rather highly degenerate states.
As seen in Fig.~\ref{fig:dosSplitting},
these states are grouped by Hund's rule coupling,
i.e., by the energy difference of the values $U_J$. The different 
clusters of energy levels are separated by well-defined gaps of about 1eV.

When spherical symmetry is broken the ${\cal J}_z$ degeneracy is lifted. This
is seen from the inset of Fig.~\ref{fig:dosSplitting}. 
It shows the evolution of the lowest 66 eigenstates
as a function of the symmetry breaking parameter 
$\Delta t=t_{3/2}-t_{1/2}$. The value of
$t_{3/2}+t_{1/2}=0.2$ is thereby kept fixed. Five different groups of
levels are noticed which for $\Delta t=0$ correspond to the values
(ordered by increasing energy)
${\cal J}_z=5/2$, 3/2, 7/2, 13/2 and a 34 fold degenerate level consisting
of ${\cal J}_z=15/2$ and 17/2 states. 
When $\Delta t\neq 0$ the ${\cal J}_z$ degenerate states
split into doublets.
For $\Delta t\simeq 0.06$ one state from the originally
34-fold degenerate states becomes lowest in energy.
This marks the transition to the ``high-spin'' state with 
${\cal J}_z=15/2$.

\subsection{Field dependence}\label{subSecField}

\begin{figure}[h t b]
\begin{center}
\includegraphics[width=0.7\columnwidth]{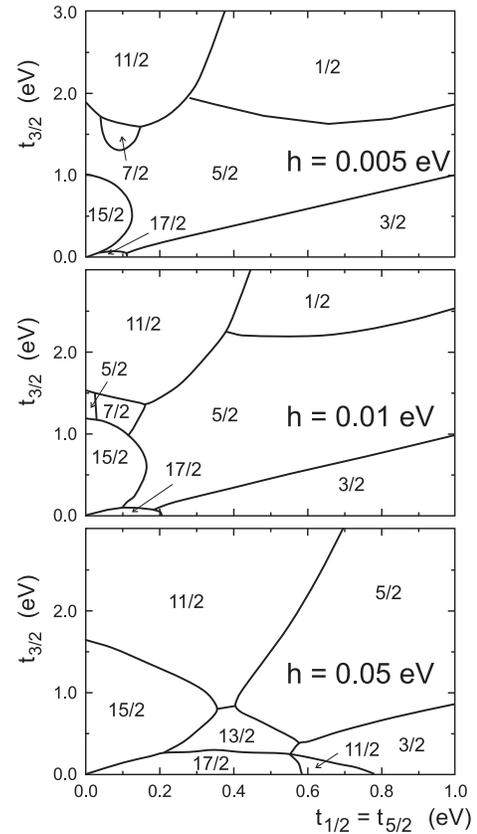}
\end{center}
\caption{Phase diagram of the two-site cluster 
in the presence of a magnetic field: 
from top to bottom  $h=$0.005,\, 0.01, and 0.05\,eV, respectively.
}\label{fig:JzRestrictedModelMagneticField}
\end{figure}

In the following we want to discuss the effect of an
internal or molecular field on the phase diagram.  
Fig.~\ref{fig:JzRestrictedModelMagneticField} shows the changes
in the phase diagram caused by a ferromagnetic field of three
different magnitudes. One notices the appearance of new ``high-spin'' as well
as ``intermediate spin'' phases \cite{Bronger}. Now the phases with low ${\cal
J}_z$ values require larger hopping matrix elements in order to be
realized. The regions of "high-spin" states grow with increasing field and for
small  values of $t_{3/2}$ a new phase with  ${\cal J}_z=17/2$ appears. This
state is similar to the ${\cal J}_z=15/2$ state and has the form
\begin{equation}
  |\Psi \rangle_{\frac{17}{2}} = 
  \frac{ c^\dagger_{1/2}(a) + c^\dagger_{1/2}(b)}{\sqrt{2}}\,
  c^\dagger_{5/2}(a) c^\dagger_{3/2}(a)
  c^\dagger_{5/2}(b) c^\dagger_{3/2}(b)\, | 0 \rangle. 
\label{eq:FerroHighState17}
\end{equation}

\section{Three sites}

In this section, we extend our analysis to a 3-site cluster.
In order to approach the valency $f^{2.7}$ 
present in UPt$_3$, 
we choose a cluster with eight electrons 
(average $f$~valency 2.67). 
This choice also avoids the Kramers degeneracy which 
was present in the 2-site cluster and which would 
again be present in the 3-site model with 7 electrons, 
i.~e., the other possible electron number with
mixed valency.
We put the sites on a ring order to 
work with a model in which all sites are equivalent.
As a consequence, 
AFM correlations are subject to frustration. 
Furthermore
the sign of the matrix elements $t_{j_z}$'s is now relevant: the 
eigenstates of the kinetic energy term in the Hamiltonian consist of a 
doublet with energy 
$t_{j_z}$ and a singlet state with energy $-2 t_{j_z}$. 
Thus, for $t_{j_z}<0$
occupancy of $j_z$ orbitals two of the three sites is favored whereas
for $t_{j_z}>0$ a single occupancy is.
Below, we consider both negative and positive hopping parameters,
assuming the same sign for all $t_{j_z}$.
All calculations are done in the subspace of only
atomic $f^2$ and $f^3$ configurations. 

\subsubsection{\large The case $t_{j_z}\!\!<0$.}

\begin{figure}[hbt]
\includegraphics[width=1.0\columnwidth]{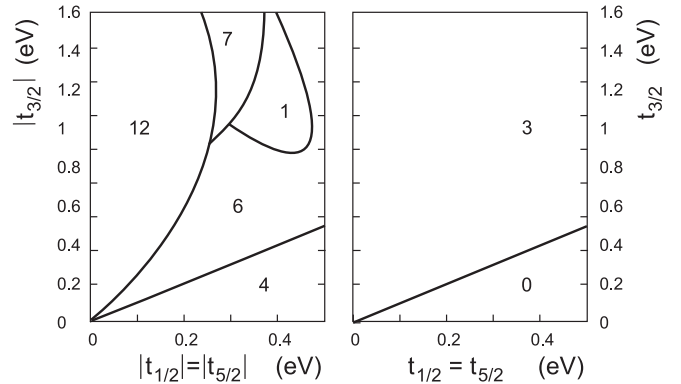}
\vspace{.3cm}
\caption{\label{3sitePM}
Phase diagram for three sites for negative (left panel) 
and positive hopping (right panel).
}\end{figure}

The left side of Fig.~\ref{3sitePM}  shows the value of the total angular momentum 
${\cal J}_z$ of the three-site cluster as a function
of negative $t_{1/2}=t_{5/2}$ and $t_{3/2}$. 
As for the two-site cluster,
a ferromagnetic high-spin phase appears in the
region $t_{1/2}/t_{3/2}\ll 1$. 
This phase has ${\cal J}_z=12$ and can be understood as
a straightforward generalization of the 
${\cal J}_z=15/2$ phase of the two-site model.
It is an eigenstate of both the Coulomb and the kinetic 
terms. Only orbitals with $j_z>0$ are occupied.
The $j_z=+5/2$ and $+1/2$ orbitals are filled 
with one electron per site and the remaining two electrons 
occupy $j_z=3/2$ orbitals. 
In contrast to this simple ferromagnetic state, all
other phases in Fig.~\ref{3sitePM}a 
have a more complicated structure.
Along the isotropic line of the phase diagram, $t_{j_z}=t$, 
the ground state has ${\cal J}=6$ for all finite values of 
$t_{\nu}$.

The phases differ also in their magnetic correlations.
Two of them have well pronounced correlations.
The ${\cal J}_z=1$ phase has antiferromagnetic 
correlations ($\sum_{\langle n m\rangle} \big< {\bf  J}(n) 
\cdot {\bf  J}(m)\big> <0$),
whereas the ${\cal J}_z= 7$ and 12 phases
have ferromagnetic correlations.  Correlations within the 
${\cal J}_z=6$ phase
are by about a factor 5 smaller than those of the ${\cal J}_z=12$ phase. 
The other two phases, ${\cal J}_z=4$,$6$ have generally small magnetic correlations,
which can be either ferro- or antiferromagnetic, depending on $t_{1/2}$ and
$t_{3/2}$.

\subsubsection{\large The case $t_{j_z}\!\!>0$.}

We now turn to the case with
positive hopping matrix elements. The ground state 
along the isotropic line has ${\cal J}=4$ for all finite values of $t_{\nu}$ .
Away from the isotropic line one finds 
${\cal J}_z=0$ for $t_{1/2}=t_{5/2}>t_{3/2}$
and ${\cal J}_z=3$ for $t_{1/2}=t_{5/2}<t_{3/2}$, 
see Fig.~\ref{3sitePM}b. 

As the kinetic energy penalizes states in which two or more 
electrons on different sites occupy the same $j_z$ orbital, 
the ferromagnetic state which was found for $t_{j_z}\!\!<0$ 
is no longer favorable.
Most of the phase diagram is dominated by AFM correlations. However,
for large anisotropies $t_{3/2}/t_{1/2}\gg 1$ these correlations
weaken continously with increasing ratio $t_{3/2}/t_{1/2}$ and
vanish near $t_{1/2}=0$.

\section{Partial localization}\label{secPartial}
In the following we want to demonstrate that strong on-site
correlations result in a considerable enhancement of anisotropies
in the bare hopping matrix elements. This can lead to a 
a localization of electrons in orbitals with relatively weak hybridization.
The latter is effectively reduced to zero in those cases.

In order to quantify the degree of localization or, alternatively, of the 
reduction of hopping of a given $j_z$ orbital by local correlations,
we calculate the ratio of the
$j_z$| projected kinetic energy $T_{j_z}$  
and the bare matrix element $t_{j_z}$,
i.~.e.,
\begin{eqnarray} 
\label{eqDefDeltaTJ}
  \lefteqn{ 
  \frac{T_{j_z}}{t_{j_z}}
= }
  \\ && 
  \sum_{\langle nm \rangle, \pm} \langle\Psi_{\mathrm gs} | 
  (  c^\dagger_{\pm j_z}(n)\,  c_{\pm j_z}(m) + h.c. ) 
|\Psi_{\mathrm gs} \rangle 
  \nonumber 
  \, . 
\end{eqnarray}
The ground-state wavefunction $| \Psi_{\mathrm gs} \rangle$ contains
the strong on-site correlations. A small ratio of $T_{j_z}/t_{j_z}$
indicates partial suppression of hopping for electrons in the
$\pm j_z$ orbitals. Two kinds of correlations may contribute to that
process. The first one is based on the reduction of charge
fluctuations to atomic $f^2$ and $f^3$ configurations. This is a result
for large values of $U_J$ and can be studied by setting all $U_J$ equal
to a value much larger that the different $t_{j_z}$. The second one is due
to differences in the $U_J$ values, i.~e.~, $U_{J=4}<U_{J=2}<U_{J=0}$.
The differences in the $U_J$ values are the basis of Hund's rules.
Hopping counteracts Hund's rule correlations and vice versa.
What we want to stress is the fact that those correlations can lead to a 
complete suppression of hopping channels except for the dominant 
one which is influenced only little.

\subsection{Two-sites cluster}

\begin{figure}[h t b]
\begin{center}
\includegraphics[width=.5\columnwidth,angle=0,clip]{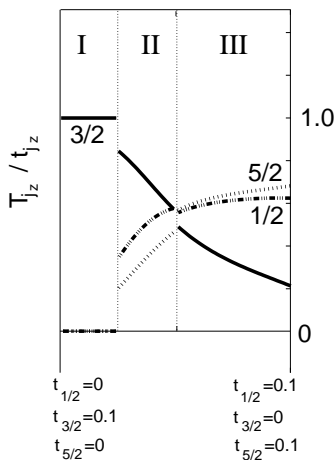}
\end{center}
\caption{
Values
 $T_{j_z}/t_{j_z}$ for the two-site cluster along the line connecting linearly
the points written below the figure. 
Phases with ${\cal J}_z=15/2$, 5/2 and 3/2, are labeled with
I, II and III respectively.
}\label{fighop1}
\end{figure}

Results for the ratios $T_{j_z}/t_{j_z}$ are shown in 
Fig.~\ref{fighop1}. For a better understanding one should consider
that figure in parallel with Fig.~\ref{f2f3_PhasenTT2.eps}. The
parameter range in Fig.~\ref{fighop1} corresponds to 
a straight line in Fig.~\ref{f2f3_PhasenTT2.eps} intersecting the 
abscissa at $t_{3/2}=0.1$ eV and the ordinate at $t_{1/2}=t_{5/2}=0.1$ eV.
This line crosses three different phases with ${\cal J}_z=15/2$, 5/2 and 3/2,
labeled I, II and III respectively. One notices that
in region I only the dominant hybridization of the $j_z=1/2$ and $j_z=5/2$
orbitals is completely suppressed. In regions II and III the correlation
effects on different orbitals are not so extremely different. In a forthcoming
paper we show how the effects of the correlations on the ratios
$T_{j_z}/t_{j_z}$ are changed when all $U_J$ values are assumed to be equal.

To understand the suppression of  
the $T_{j_z}$ in the limit of very small $t_{j_z}$ it is sufficient to
treat the hopping perturbatively, this is discussed in Appendix
\ref{appPerturbation}.

\subsection{Three sites}

In the case of a three-sites cluster the sign of the matrix
elements $t_{j_z}$ matters because of the respective degeneracy of the
ground state.
We begin with $t_{j_z}<0$. 
 
\begin{figure}[h t b]
\begin{center}
\includegraphics[width=.5\columnwidth,angle=0,clip]{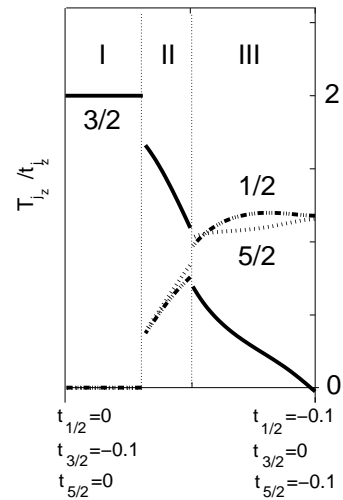}
\end{center}
\caption{
Values $T_{j_z}/t_{j_z}$ of the three-sites model with $t_{j_z}<0$
along the line connecting linearly the $t_{j_z}$ values indicated in the
figure. 
}\label{fighop3}
\end{figure}

Shown in Fig.~\ref{fighop3} are results for different parameter
choices. Again, for a better understanding, one should consider in parallel
the phase diagram in the left hand side of Fig.~\ref{3sitePM}.
Again, three different phases are involved and they are labeled as before
by I, II and III in Fig.~\ref{fighop3}.
Altogether, the figure looks similar to Fig.~\ref{fighop1}. There is again
a regime of total suppression of the $t_{1/2}$ and $t_{5/2}$ matrix elements
(region I)  and also in region III the smaller of the $t_{j_z}$, i.~e.,
$t_{3/2}$ is strongly reduced to an effective $T_{3/2}$. The case
of equal Coulomb integrals $U_J$, i.~e., $\Delta U_J=0$ is discussed in an
extended version of this paper.

\begin{figure}[h t b]
\begin{center}
\includegraphics[width=.5\columnwidth,angle=0,clip]{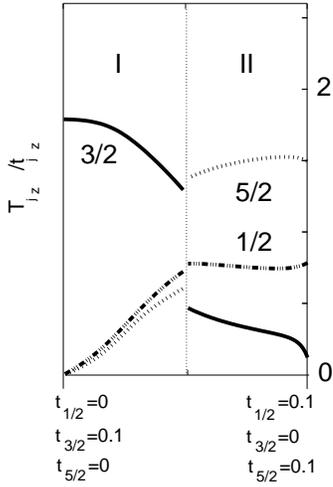}
\end{center}
\caption{
Values $T_{j_z}/t_{j_z}$ of the three-sites model with $t_{j_z}>0$
along the  line connecting the $t_{j_z}$ values written
below the figures. 
}\label{fighop2}
\end{figure}

Next we turn to the case of positive $t_{j_z}>0$. Corresponding results
are shown in Fig.~\ref{fighop2} which should be considered in connection
with the right hand side of Fig.~\ref{3sitePM}. There it is seen that
two different phases with ${\cal J}_z=3$ and 0 are involved with the chosen
parameters $t_{j_z}$. In distinction to Figs.~\ref{fighop1} and \ref{fighop3} there
is no regime of total suppression of hopping matrix elements. But
a strong enhancement of anisotropies by the correlations is noticeable
also here. Again, the overall features are very similar to those in
Figs.~\ref{fighop1} and \ref{fighop3} except that the intermediate regime is
missing here. This is of little surprise, since we are dealing with 
local correlations so that the effects should be insensitive to the
precise form of the chemical environment.

It is instructive to take a closer look at the ground state.
For $t_{1/2}/t_{3/2}<1$, the ground state is twofold degenerate
with ${\cal J}_z=\pm3$ (see Fig.~\ref{3sitePM}). 
When $t_{1/2}=t_{5/2}$ are small,
$t_{1/2}/t_{3/2}\ll 1$,
the $f^2$ configurations are strongly dominated
by ones in which the orbitals
$j_z=\pm 1/2$ and $j_z=\pm 5/2$ are occupied with a ferromagnetic
alignment of the electrons.
The admixture of other states is less then 1.5 \% of 
the norm for $t_{3/2}=0.5$ and $t_{1/2}=t_{5/2}=0.05$.
Furthermore, the wavefunction is strongly dominated, in fact to
87 \%, by configurations in which the $j_z=\pm 1/2$, $\pm 5/2$ orbitals 
form
an Ising-like frustrated AFM background
for mobile $j_z=\pm 3/2$ orbitals.
We describe this AFM background by states
\begin{eqnarray}
  \lefteqn{ 
  | \downarrow \uparrow \uparrow \rangle :=  
  \Big( c_{-5/2}^\dagger(a) \,c_{-1/2}^\dagger(a) \Big) 
  } 
  \\ && 
  \Big( c_{ 1/2}^\dagger(b) \,c_{ 5/2}^\dagger(b) \Big)
  \Big( c_{ 1/2}^\dagger(c) \,c_{ 5/2}^\dagger(c) \Big) 
  \,\, \Big| 0 \rangle \nonumber
\end{eqnarray}
($   | \uparrow \downarrow \uparrow \rangle$ 
and $| \uparrow \uparrow \downarrow \rangle$ 
are defined analogously).

We further introduce operators
$ {\cal C}^\dagger$ for itinerant 
electrons in the ${\pm3/2}$ orbitals
\begin{equation}
  {\cal C}_{\pm3/2}^\dagger =
   \frac{\alpha_{\pm}\,c_{\pm 3/2}^\dagger(a)+
   c_{\pm 3/2}^\dagger(b)+c_{\pm 3/2}^\dagger(c)}
   {\sqrt{2+\alpha_{\pm}^{\,2}}} 
\end{equation} 
with a number $\alpha_\pm$ of order 1 
taking into account that due to the background 
not all three sites are equivalent. 
The 
ground-state wavefunction (with ${\cal J}_z=3$) is then of the
approximate form
$$|\Psi_{{\cal J}_z=3} \rangle\simeq 
\left(|h_0;\downarrow \uparrow   \uparrow\rangle
    + |h_0;\uparrow   \downarrow \uparrow \rangle +
    + |h_0;\uparrow   \uparrow   \downarrow\rangle \right)/\sqrt{3}$$ 
where
\begin{eqnarray}
  |h_0;\downarrow \uparrow \uparrow \rangle & = & 
  \frac{1}{\sqrt{20}} \hat {\cal P}_{f^2-f^3}\,
   {\cal C}_{+3/2}^\dagger  {\cal C}_{-3/2}^\dagger 
  | \downarrow\uparrow\uparrow \rangle 
 \,.
 \label{Gutzwiller3}
\end{eqnarray}
The projector $\hat{\cal P}_{f^2-f^3}$ projects onto local
$f^2$ and $f^3$ configurations.
Furthermore 
$|h_0;\uparrow\downarrow \uparrow \rangle$, 
$|h_0;\uparrow \uparrow\downarrow\rangle$
are obtained from $|h_0;\downarrow \uparrow \uparrow\rangle$
via cyclic
permutation of sites. 
Note that the kinetic energy contributions
$T_{1/2}$ and $T_{5/2}$ calculated with
the wavefunction~(\ref{Gutzwiller3}) vanishes exactly.

As the ground state can be written approximately as a sum of simple products 
of itinerant $j_z=\pm 3/2$ particles 
and a magnetic background,
one expects to find a similar form for the lowest excited state.
This is indeed the case:
The first excited state has again  
${\cal J}_z=\pm 3$ but with a fourfold rather than twofold
degeneracy. The excited states can be very well approximated by a 
difference of
any two of the three states $|h_0;\downarrow \uparrow \uparrow\rangle$,
 $|h_0;\uparrow\downarrow \uparrow \rangle$, 
$|h_0;\uparrow \uparrow\downarrow\rangle$, e.~g.
$\left( 
|h_0;\downarrow \uparrow\uparrow \rangle-
|h_0;\uparrow\downarrow \uparrow \rangle
\right)\sqrt{2}$.
For fixed ${\cal J}_z=3$ or   ${\cal J}_z=-3$ these states span 
a subspace of dimension two. 
We may view the first excited state
as an excitation of the frustrated AFM background which
is dominated by the $j_z=\pm 5/2$,$\pm1/2$ orbitals, while
the electrons/holes in the $j_z=\pm 3/2$
orbitals simply adjust to  the slow dynamical changes of
the magnetic background.

The magnetic correlations of the ground state 
are also interesting.
The overall magnetic correlations in 
$|\Psi_{{\cal J}_z=3}\rangle$ are weakly ferromagnetic, i.~e.,
$\langle \Psi_{{\cal J}_z=\pm3} | \sum_{<m,n>} 
        {\bf J}(m) 
  \cdot {\bf J}(n) 
  | \Psi_{{\cal J}_z=\pm 3} \rangle=3$. 
But because of
$ \langle \Psi_{{\cal J}_z=3} | \sum_{<m,n>} 
       {\bf J}(m) 
  \cdot {\bf J}(n) 
  | \Psi_{{\cal J}_z=-3} \rangle=-3$
one can construct states 
 $  |\Psi_{{\cal J}_z=-3} \rangle\pm |\Psi_{{\cal J}_z=3} \rangle$ 
which have no magnetic correlations. 
Thus, we have a situation with a localization 
mechanism akin to a Mott transition, 
but without the magnetic correlations 
which are usually present in Mott systems.

\section{Discussions and Summary.}

In this investigation we have studied the effects of strong intra-atomic 
correlations on anisotropies in the hybridization of different 
5$f$ orbitals with the 
environment. For that purpose we considered two-site and 
three-site clusters which allow for a complete numerical diagonalization 
of the corresponding Hamiltonian. We demonstrated that in 
particular Hund's rule correlations  enhanced  
strongly anisotropies in the hybridizations. For a certain range of parameters 
this may result in a complete suppression of the effective hybridization 
except for the largest one, which remains almost unaffected. This provides 
for a microscopic picture of partial localization of 5$f$ electrons, 
a phenomena, observed in a number of experiments on U compounds. In fact, 
previous work on the de Haas-van Alphen frequencies and anisotropic 
effective masses in UPt$_3$ and UPt$_2$Al$_3$ could explain the experimental 
data very well by assuming, that two of the 5$f$ electrons of the U ions 
remain localized while a third one is itinerant \cite{GYF02,GEZJapan}. 
The latter is in an orbital with $j_z=3/2$ in both cases, and a $j$-$j$ 
coupling scheme is used. Previous work on the molecule 
uranocene U(C$_8$H$_8$)$_2$ gave a hint on  partial localization \cite{KLDF},
but the present work provides a detailed justification for such a feature.    
By including an external field we obtained a rich magnetic phase diagram. 
We plan to investigate its relevance to compound such as 
UGe$_2$. We note that intra-atomic excitations within the system of two 
localized 5$f$ electrons might provide the required bosonic excitations 
for itinerant electron attraction and Cooper-pair formation
\cite{Fulde70,Sato01,Hotta}. One may also speculate 
about the application of the present findings to multiband Hubbard models. 
They are used to describe correlations in transition metal compounds 
\cite{chapter11}. 
This problem  is
left for a future investigation as well 
as the one of a U impurity embedded in a weakly correlated chemical 
environment. Also a molecular-field description of the effects found here,
 e.g., by introducing auxiliary fields to model the correlation effects 
is an interesting subject  
of future research. In an extended version of this paper we plan to discuss 
also the phenomenon of charge disproportionation, which may occur in clusters
 of the type investigated here.

\begin{appendix}  
\section{Perturbation theory}\label{appPerturbation}

In the limit $t_{j_z}\rightarrow 0$, the 180 states of 
products of local $f^2$ configurations $|4,J_z'\rangle$
and local $f^3$ configurations $|9/2,J_z\rangle$
are degenerate. 
The corresponding states are direct products of
maximal local $\cal J$ states, i.e.,
three-particle states with $J(a)(J(b))=9/2$ 
and two-particle states
with $J(b)(J(a))=4$.

Finite hopping  splits the
manifold in first order degenerate perturbation theory. 
The transition matrix element for hopping with initial product state
$|f^3;\frac{9}{2},J'_z   \rangle$
$|f^2;          4,J_z\rangle$
and final state 
$|f^2;          4,J_z'-j_z \rangle$
$|f^3;\frac{9}{2},J_z +j_z \rangle$
is 
\begin{eqnarray}\label{eq:A1}
  \sum_{j_z} t_{j_z} 
 && { \textstyle \langle f^3;\frac{9}{2}, J_z +j_z    | 
    c^\dagger_{j_z} |  f^2;          4, J_z  \rangle } \nonumber 
\\ && \times 
  { \textstyle \langle f^2;          4,      J_z^{\prime }-j_z | 
    c_{j_z} |  f^3;\frac{9}{2} J_z^{\prime } \rangle } 
   \nonumber \\
  =   \frac{33}{14} &&\sum_{j_z} \cdot  t_{j_z}  
\displaystyle C_{5/2, j_z; 4, J_z}^{9/2, J_z +j_z}  
	      (C_{5/2, j_z; 4, J_z'-j_z}^{9/2, J_z'})^* 
\end{eqnarray}

\noindent where the $C^{a, b}_{c, d}$ are Clebsch-Gordan coefficients.

In the plane spanned by $t_{3/2}$ and $t_{1/2}=t_{5/2}$, for small 
$t_{j_z}$
transitions occur near $t_{1/2}/t_{3/2}\simeq 0.35$ and at
the isotropic line,
see Fig.~\ref{f2f3_PhasenTT2.eps}. 
These transitions are reproduced 
with the perturbative treatment. Furthermore,
the perturbative analysis also describes the degeneracies of the
ground states correctly. 

Of particular interest is the behavior along the isotropic line $t_{5/2} =
t_{3/2} = t_{1/2} = t$ where the lifting of degeneracy can be solved in closed
form. In this limit {\boldmath${\cal J}$}$^2$ provides an additional good quantum number. We
therefore consider basis states which are eigenstates of {\boldmath${\cal
J}$}$^2$, i.e., $| 9/2~4{\cal J}{\cal J}_z \rangle$ and $|4~ 9/2~ {\cal J}{\cal
J}_z \rangle$. They are obtained as linear combinations of the above-mentioned
product states $|f^3; 9/2, {\cal J}_z - J_z \rangle | f^2; 4, J_z \rangle$ and
$|f^2; 4, {\cal J}_z - J_z - j_z \rangle | f^3; 9/2, J_z + j_z \rangle$. In
this basis, the hopping matrix elements depend only upon ${\cal J}$ and can be
expressed in terms of the 6-$j$ symbols 

\begin{equation}
t \langle \frac{9}{2}~4~ {\cal J} {\cal J}_z \mid \sum_{j_z} c^\dagger_{j_z}
(a) c_{j_z} (b) \mid 4~\frac{9}{2} {\cal J} {\cal J}_z \rangle = \frac{33}{14}
t~ \tau({\cal J}) 
\end{equation} 
\noindent where
\begin{equation}
\tau ({\cal J}) = (-1)^{\cal J}~~10~~\left( 
\begin{array}{ccc}
4 & \frac{5}{2} & \frac{9}{2}\\[1ex]
4 & {\cal J} & \frac{9}{2}
\end{array} \right)
\end{equation}
 
\noindent accounts for the variation with ${\cal J}$ of the hopping matrix
element. 

This result allows for a simple interpretation. In the $f^2 - f^3$ model, the
total angular momentum ${\cal J}$ of the 5-electron states $| 9/2~~4~{\cal
J}~{\cal J}_z \rangle$ and $| 4~~9/2~{\cal J}~{\cal J}_z \rangle$ arises from
the coupling of three angular momenta. They
characterize $f^2$ configurations at the two sites $a$ and $b$ with angular
momenta 4 and an
additional electron which can sit at either site. The angular momentum $j =
5/2$ of this electron couples with that of the corresponding
$f^2$-configurations yielding the angular momentum $J(a) = 9/2$ or $J(b) = 9/2$
of the $f^3$-configurations at site $a$ or $b$, respectively. Technically
speaking, the states $| 9/2~~4~{\cal J}~{\cal J}_z \rangle$ and $| 4~~9/2~{\cal
J}~{\cal J}_z \rangle$ arise from applying different coupling schemes to the
three above-mentional angular momenta. Hopping between the two sites connects
the two types of states. As a consequence, the hopping matrix elements can be
expressed in terms of the 6$j$-symbols which transform between the two
different coupling schemes. 

The variation with ${\cal J}$ of the hopping matrix elements is easily
evaluated yielding 

\begin{equation}
\tau({\cal J}) = (1 + 2 {\cal J}) \frac{183681-8424 {\cal J} - 8344 {\cal J}^2
+ 160 {\cal J}^3 + 80 {\cal J}^4}{1064448}
\end{equation} 

\begin{figure}[hbt]
\begin{center}
\includegraphics[width=0.8\columnwidth]{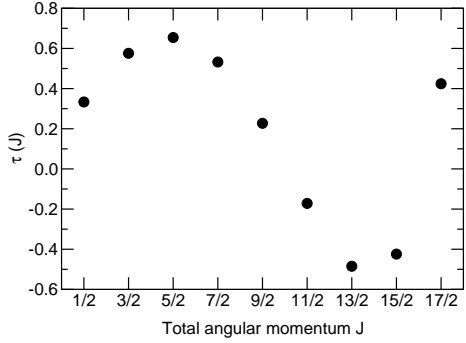}
\end{center}
\caption{Variation of $\tau({\cal J})$ with ${\cal J}$ in the isotropic
case.\label{fig:6j} 
}\end{figure}

\noindent Here ${\cal J}$ is restricted to half-integer values $1/2 \leq {\cal
J} \leq 17/2$. The variation with ${\cal J}$ of $\tau({\cal J})$ is displayed
in Fig.~\ref{fig:6j}. The maximum value for $|\tau({\cal J})|$ is obtained for
${\cal J} = 5/2$ corresponding to a sixfold degenerate ground state in the
isotropic case.


\section{Analysis of ground states}\label{appAnalysis}

\begin{figure}[hbt]
\begin{center}
\includegraphics[width=0.8\columnwidth]{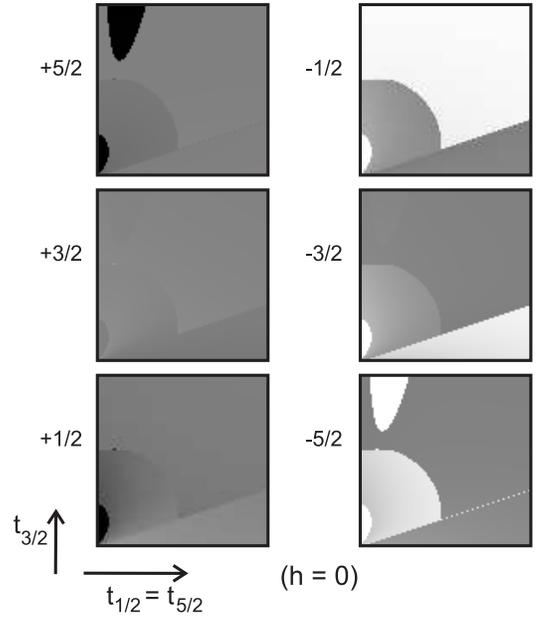}
\end{center}
\vspace*{-.2cm}
\caption{\label{figWWTT}
Gray-scale plot of orbital occupations as function of hopping matrix elements
for $j_z=+\frac{5}{2}, ... , -\frac{5}{2}$ 
in the presence of an infinitesimal ferromagnetic field. 
Black: fully occupied; white: empty.
}\end{figure}

At the phase boundaries in Fig.~\ref{f2f3_PhasenTT2.eps} the jumps of the $z$
component of the total magnetization of a two-site cluster coincide with
changes of the orbital-resolved occupation number $n _{j_z}(m)$ in the
ground-state wavefunction. There are shown as gray-scale plots in
Fig.~\ref{figWWTT}. They confirm the scenario developed in the body of the text
and allow for some additional observations. The figure corresponds to the
$t_{3/2} - t_{1/2}$ phase diagram shown in Fig.~\ref{f2f3_PhasenTT2.eps} in an
infinitesimal ferromagnetic field. For weak hopping with $t_{3/2}\gg
t_{1/2}=t_{5/2}$, the $j_z=+5/2$ and the $j_z=1/2$ orbitals are fully occupied
on both sites, whereas $j_z=+3/2$ is half-occupied, allowing for a maximal
kinetic energy gain. The corresponding total angular momentum component ${\cal
J}_z=2\cdot\frac{5}{2}+2\cdot\frac{1}{2}+\frac{3}{2}=15/2$. Similarly for large
$t_{3/2}$ values the ${\cal J}_z=11/2$ phase is formed in which one of the
$j_z=1/2$ electrons is promoted to $j_z=-3/2$. 

In all other phases, the average occupations of all orbitals are close to one
half or zero. Occupations near one half could indicate frequent hopping. Such a
'paramagnetic' interpretation is appropriate only if $t_\nu$ is much larger
than the energy differences $\Delta U_J$ resulting in Hund's rule. The three
phases to be discussed are best classified by looking at the particular $j_z$
orbital which remains unoccupied (light-colored areas in the
figure). Obviously, for the case $t_{3/2} < t_{1/2}=t_{5/2}$, the loss of
hopping energy gain is minimal if $j_z=-3/2$ remains empty, yielding ${\cal
J}_z=3/2$. In the case $t_{3/2} > t_{1/2}=t_{5/2}$, either the orbital
$j_z=-1/2$ or $j_z=-5/2$ is unoccupied, leading to ${\cal J}_z=+1/2$ and ${\cal
J}_z=+5/2$, respectively. It should be kept in mind that the Coulomb
interaction mixes various $j_z$ orbitals. The concept of 'occupying' $j_z$
states is therefore only approximate. This is reflected in small changes of the
occupation within each phase.  

\end{appendix}

\end{multicols}
\end{document}